\documentclass[aps,showpacs, nofootinbib]{revtex4}

\input amssym.tex
\input amssym.def

\newcommand{\be}{\begin{equation}}
\newcommand{\ee}{\end{equation}}
\newcommand{\ben}{\begin{eqnarray}}
\newcommand{\een}{\end{eqnarray}}

\newcommand{\la}{{\lambda}}

\newcommand{\cO}{{\cal O}}

\newcommand{\p}{\partial}
\newcommand{\na}{\nabla}

\newcommand{\tchi}{\tilde \chi}

\newcommand{\tim}{{\tilde \mu}}
\newcommand{\tom}{{\tilde \omega}}

\newcommand{\Dsl}{{\slash \negthinspace \negthinspace \negthinspace \negthinspace  D}}
\newcommand{\tM}{{\tilde M}}
\newcommand{\tm}{{\tilde m}}

\newcommand{\tV}{{\tilde V}}
\newcommand{\tOm}{{\tilde \Omega}}

\newcommand{\tS}{\tilde S}

\newcommand{\ep}{\epsilon}

\newcommand{\ga}{\gamma}

\pacs{04.50.+h}

\begin{document}

\title{Dirac Fermions in Non-trivial Topology Black Hole Backgrounds}

\author{Marek G\'o\'zd\'z}
\email{mgozdz@kft.umcs.lublin.pl}
\affiliation{
Institute of Informatics, Maria Curie-Sk{\l}odowska University, \\
20-031 Lublin, pl.~Marii Curie-Sklodowskiej 5, Poland}

\author{{\L}ukasz Nakonieczny and Marek Rogatko}
\email{rogat@kft.umcs.lublin.pl, 
marek.rogatko@poczta.umcs.lublin.pl}
\affiliation{Institute of Physics \protect \\
Maria Curie-Sklodowska University \protect \\
20-031 Lublin, pl.~Marii Curie-Sklodowskiej 1, Poland }


\date{\today}

\begin{abstract}
We discuss the behaviour of the Dirac fermions in a general spherically symmetric black hole
background with a non-trivial topology of the event horizon. 
Both massive and massless cases are taken into account.
The analytical studies of
intermediate and late-time behaviour of massive 
Dirac hair in the background of a black hole with a global monopole and 
dilaton black hole pierced by a cosmic string will be conducted.
It was revealed that in the case of a global monopole swallowed by a static black hole
the intermediate late-time behaviour depends on the mass of the Dirac field,
the multiple number of the wave mode and the global monopole parameter.
The late-time behaviour is quite independent of these factors and has the decay rate proportional to $t^{-5/6}$.
As far as the black hole pierced by a cosmic string is concerned the intermediate late-time behaviour 
depends only on the hair mass and the multipole number of the wave mode
while the late-time behaviour dependence is the same as in the previous case.
The main modification stems from the topology of the $S^2$ sphere pierced by a cosmic string. This factor
modifies the eigenvalues of the Dirac operator acting on the transverse manifold.

\end{abstract}

\maketitle

\section{Introduction}
\par
During past decades the study of fermions behaviour in various backgrounds has attracted
much interests. 
Exact solution of Dirac equation in curved spacetime is a very useful tool
in investigations of physical properties of particles in adequate spacetimes.
As far as black holes is concerned, they are in this category, because of the fact that
their better understanding acquires a detailed study of various matter fields in its vicinity
(see \cite{chandra} and references therein).
Dirac fermion fields were studied in the context of Einstein-Yang-Mills background \cite{gib93}
found by Bartnik and McKinnon \cite{bar88}. 
Fermion fields were analyzed in the near horizon limit of an extreme Kerr black hole \cite{sak04}.
It turned out that the extreme Reissner-Nordstr\"om (RN) case  \cite{loh84} having both magnetic and electric charges
was destroyed in the presence of a massless fermion field.
It was also shown \cite{fin00}, generalizing previous results \cite{findir}, that 
the only black hole solution of four-spinor Einstein-dilaton-Yang-Mills equations
were those for which the spinors vanished identically outside black hole. It physically means that
the Dirac particles either enter the black hole or escape to infinity.
In Refs.\cite{br1,br2} Dirac fields were considered in Bertotti-Robinson spacetime.
On the other hand, Dirac fields as a cosmological solution with
a homogeneous Yang-Mills fields acting as an energy source were analyzed in Ref.\cite{gib94}.
Dirac equation, via the Newman-Penrose formalism, in the context of Robertson-Walker spacetime was considered
in \cite{zec95}. Recently, the behaviour of massive Dirac fields on the background of a charged
de Sitter black holes was elaborated in Ref.\cite{bel09}.
\par
Regardless of the details of the gravitational collapse, the resultant black hole is described only by few parameters
such as mass, charge and angular momentum. It is the famous Wheeler dictum that {\it black holes have no hair}.
But the tantalizing question is how the loss of black hole hair takes place.
The first contributions towards this problem were given in Refs.\cite{pri72,gun94}, where 
it was shown that the late-time behavior is dominated by the factor $t^{-(2l + 3)}$, for each 
multipole moment $l$ and the decay-rate 
along null infinity and along the future event horizon was governed
by the power laws $u^{-(l + 2)}$ and $v^{-(l + 3)}$, where $u$ and $v$ were
the outgoing Eddington-Finkelstein (ED) and ingoing ED coordinates.
In Refs.\cite{pir1}-\cite{pir3} charged scalar hair decay cases were considered.
It turns out that a charged scalar hair
decayed slower than a neutral one.
The nearly extreme RN spacetime was treated in Ref.\cite{ja} where it was
shown that the inverse power law behaviour of the dominant asymptotic tail is of the form
$t^{-5/6} \sin (m t)$, being independent of $l$. The asymptotic tail behaviour 
of a massive scalar field
was also studied in Schwarzschild spacetime \cite{ja1}. The oscillatory
tail of the scalar field has the decay rate of $t^{-5/6}$ at                                   
asymptotically late time.
The power-law tails in the evolution of a charged massless and self-interacting scalar field around a fixed
background of dilaton black hole was studied in Refs.\cite{mod01a,mod01b}, where
the inverse power-law relaxation of the fields at future timelike infinity, future null infinity and
along the outer horizon of the considered black hole was found.
It was also envisaged numerically that at very late times
the oscillatory tail decay is of the form $t^{-5/6}$. These results were confirmed analytically in the 
theory with arbitrary coupling constant between dilaton and $U(1)$-gauge fields \cite{rog07}.
\par
Much attention was also paid to 
the late-time behaviour of massive Dirac fields. They were studied
in the spacetime of Schwarzschild black hole \cite{jin04}
as well as in RN black hole background \cite{jin05}. 
The stationary axisymmetric black hole case 
was studied numerically in Ref.\cite{bur04} and it was found that in the case of Kerr black hole 
the oscillatory inverse-power law of 
the dominant asymptotic tail behaviour was approximately depicted by the relation $t^{-5/6}\sin(mt)$.
In Ref.\cite{xhe06} the decay of a charged massive
Dirac hair in the background of Kerr-Newmann black hole was investigated.
It turned out that 
the intermediate late-time behaviour of the fields under consideration was dominated by an
inverse power-law decaying tail without any oscillation.
\par
In Ref.\cite{gibrog08} the analytical studies concerning the intermediate and late-time
decay pattern of massive Dirac hair on the dilaton black hole were conducted.
Dilaton black hole constitutes a static spherically symmetric solution of the theory being
the low-energy limit of the string theory with arbitrary coupling constant $\alpha$.\\
Massive vector field obeying the Proca equation of motion in the background
of Schwarzschild black hole was studied in \cite{kon07}. It was revealed that 
at intermediate late times, three functions characterizing
the field have different decay law depending on the multipole number $l$. On the contrary,
the late-time behaviour is independent on $l$, i.e., the late-time decay law is
proportional to $t^{-5/6}\sin(mt)$.
\par
Recently, there has been renewed interest in brane models in which our Universe
is represented as $(3 + 1)$-dimensional submanifold living in higher dimensional
spacetime as well as brane black holes. The decay of massive Dirac 
hair on a brane black hole was considered in \cite{br08}.
\par
It might happened that at the beginning of our universe several phase transitions lead to the 
topological defect formations \cite{vil}. The interaction of topological defects such as global monopoles or cosmic strings
with black holes has very interesting features. For instance,
the black hole global monopole system has an unusual topological property of possessing a solid deficit angle.
The physical characteristics of the above system were widely studied in literature \cite{mon}.
The decay of massive scalar hair in the background of a Schwarzschild black hole with global monopole 
was studied in Ref.\cite{hyu02}.
It happened that the topological defects makes the massive scalar field hair decay faster in the 
intermediate regime comparing to the decay of such hair on the black holes without defects. On the other hand,
the late-time behaviour was unaffected by the presence of global monopole. The Schwarzschild black hole global 
monopole system coupled to scalar fields was elaborated in Ref.\cite{che05}.
\par
On the other hand,
cosmic strings and cosmic string black hole
systems also acquire much interest.
Assuming a distributional mass source the metric of this system was derived in \cite{ary86} (the so-called
{\it thin string limit}). Later, it was revealed that it constituted the limit of much more realistic situation 
when a black hole was pierced by a Nielsen-Olesen vortex \cite{greg}. 
It turned out that for some range of black hole parameters extremal black hole
expelled the vortex (the so called {\it Meissner effect}). 
Moreover, extremal black holes in dilaton 
gravity always expel Higgs fields from their interiors \cite{mod99}.
\par
It will be not amiss to study a much more realistic case than the scalar fields, i.e., the 
behaviour of fermion fields on the background of
black holes with
non-trivial topologies of their event horizons caused, e.g., 
by topological defects and the influence of the topological defects'
parameters on the intermediate and late-time decay of massive fermion hair.\\
The main aim of our work is to elaborate the solution of the Dirac equation in the background of a general spherically
symmetric black hole spacetime. We shall try to generalize in some way the attitude presented in \cite{gib93} studying
zero modes and $ k > 0$ modes of massless and massive Dirac equation in topology non-trivial spacetimes.
First, we shall take into account the case of a Dirac equation supplemented by the Yang-Mills term. We assume the complete separation
of the degrees of freedom of the considered fields. In our work, we shall consider the case
of a black hole global monopole system which implies
\be
ds^2 = - \bigg(1 - {2 G M  \over r} \bigg)dt^2 +
{dr^2 \over \bigg(1 - {2 G M  \over r} \bigg)} + b^2~ r^2 d\Omega^2,
\label{monopol}
\ee
as well as the black hole pierced by a cosmic string,
which line element in the {\it thin string} approximation of the Nielsen-Olesen vortex passing through 
the spherically symmetric static dilaton black hole. We shall consider the dilaton gravity with an arbitrary
coupling constant $\alpha$. Namely, the line element in question yields
\be
ds^2 = - \bigg( 1 - {r_{+} \over r}
 \bigg)\bigg( 1 - {r_{-} \over r}
 \bigg)^{{1 - \alpha^2} \over 1 + \alpha^2} dt^2
+ {dr^2 \over \bigg( 1 - {r_{+} \over r} \bigg)
\bigg( 1 - {r_{-} \over r} \bigg)^{{1 - \alpha^2} \over 1 + \alpha^2}}
+ R^2(r) d \tOm^2,
\label{dila}
\ee
where $R^2(r) = r^2 \bigg( 1 - {r_{-} \over r} \bigg)^{2 \alpha^2 \over 1 + \alpha^2}$
and the {\it transverse} two-dimensional manifold is given by
$d \tOm^2 = d \theta^2 + B^2 \sin^2 \theta d \phi^2$. The parameter $B$ is related to the linear mass density
of a cosmic string passing through the dilaton black hole. Of course, for dilaton black hole without
a cosmic string one should have $B = 1$.
$r_{+}$ and $r_{-}$ are the outer and the inner event horizons of the black hole. They
are related to mass $M$ and electric 
charge $Q$ of the black hole in the following way:
\be
e^{- 2 \alpha \phi} = \bigg( 1 - {r_{-} \over r} \bigg)^{2 \alpha^2 \over 1 + \alpha^2}, \qquad
2 M = r_{+} + {1 - \alpha^2 \over 1 + \alpha^2} r_{-}, \qquad
Q^2 = {r_{-}~r_{+} \over 1 + \alpha^2}.
\ee
Then, we proceed to analytical discussion of the decay of massive fermion hair on the
aforementioned backgrounds.
\par
The layout of our paper is as follows. In Sec.II we first briefly review the behaviour of Dirac fermions in the
background of a general spherically symmetric black hole. We analyze the 
massless case, zero modes of Dirac equation
in the near-horizon limit for both extremal and nonextremal black holes with
non-trivial topology of the event horizon. Then, we take into account the Dirac fermion fields for
$k > 0$ and show that the equations in question can be decoupled to the system
of second order differential equations. In the next subsection we treat the massive case of Dirac
fermions. It also happens that their equations decouple to the system of second order differential ones
with the so-called {\it supersymmetric} potentials.
In Sec.III we gave the analytic arguments concerning
the decay of Dirac massive hair in the backgrounds of topology non-trivial black holes. 
Finally, we conclude our investigations in Sec.IV.

\section{Fermions in the General Spherical Background}
In this section we shall begin our analysis by
considering the behaviour of Dirac fermion fields in a general spherically symmetric 
black hole background. The metric corresponding to the aforementioned spacetime will be given by the expression
\be
ds^2 = - A(r)^2 dt^2 + B(r)^2 dr^2 + C(r)^2~d\Omega^2,
\ee
where all the metric functions have the $r$-coordinate dependence. The {\it transverse}
metric $d\Omega^2$ depends neither on $r$-coordinate nor on $t$.
In this background, we refine our studies to the solution of the massless Dirac equation 
with Yang-Mills potential provided by
\be
i~\gamma^{\mu}(\na_{\mu}-i\lambda H_{\mu} )\psi=0,
\label{dirac}
\ee
where $\na_{\mu}$ is the covariant derivative,
$\na_{\mu}=\p_{\mu} + {1 /2}\omega_{\mu}{}^{ab} \gamma_{a}\gamma_{b}$
and $\omega_{\mu}{}^{ab}$ are the connection one-forms.
The Dirac matrices satisfy $\{ \gamma^{a},\gamma^{b}\}=-2\eta^{ab}$. They are given in the explicit form 
as follows:
\be
\gamma^{0}=
\pmatrix{ 
0 & I \cr I & 0 },
\qquad
\gamma^{a}=
\pmatrix{
0 & \sigma^{a} \cr - \sigma^{a} & 0},
\ee
where $I$ stands for the identity matrix while
$\sigma^{a}$ are the Pauli spin matrices. The vector potential of the Yang-Mills type appearing in Eq.(\ref{dirac})
will be chosen as
\be
H_{i} = {a (r) \over 2 \lambda r} \ep_{ijk}~ n^{j}~ \tau^{k},
\ee
where $\la$ is the coupling constant, $\tau^{k}$ is the generator of the group $SU(2)$ and $n^{j}$ the unit normal vector.
The spinor $\psi$ can be decomposed into its left and right chiral component, i.e.,
$
\psi =  \pmatrix{
\psi_{L} \cr \psi_{R}}.
$
It is well known (see, e.g., \cite{qft}) that for massless case the two chiralities decouple.
It enables one to rewrite the underlying Dirac equations in the form as
\ben \label{rr}
\p_{t}\psi_{R} -i \lambda~  A(r)~{\sigma}^{\mu}{H}_{\mu}~\psi_{R}
+ {A(r) \over C(r)}~\Dsl \psi_{R}+
{{\sigma}^{\mu} n_{\mu} \over C(r)}~{A(r)^{1 \over 2} \over B(r)}
\p_{r}(C(r)~A(r)^{1 \over 2}~\psi_{R}) = 0, \\
\p_{t}\psi_{L} +i\lambda ~A(r)~{\sigma}^{\mu}{H}_{\mu}~\psi_{L}
- {A(r) \over C(r)}~ \Dsl \psi_{L}-
{{\sigma}^{\mu}{n}_{\mu} \over C(r)}~ {A(r)^{1 \over 2} \over B(r)}\p_{r}(C(r)~A(r)^{1 \over 2}~\psi_{L})=0,
\een
where $\Dsl$ is the Dirac operator on the {\it transverse} manifold.
\subsection{Zero modes of the Dirac equation}
First of all we shall consider the s-wave case.
Because of the fact that we are looking for the massless spinor solution we can restrict our attention
to the case when $\psi = \psi_{R}$, without loss of generality. In our considerations we 
shall use {\it the hedgehog} spinor ansatz (see e.g.,
\cite{kim93,cho75}) in which the spinor function
$\psi$ will be spanned by two states $\chi_{1} = {1 \over 2 \sqrt{2}}
\bigg[ \pmatrix{ 1 \cr 0}_{S} \pmatrix{0 \cr 1}_{T} - \pmatrix{0 \cr 1}_{S} \pmatrix{1 \cr 0}_{T}
\bigg]$ and $\chi_{2} = \sigma_{i} n^{i}~\chi_{1}$. Moreover, it will have the property
$( \sigma_{k} + \tau_{k} ) \chi_{1} = 0$.
Consequently, the considered Dirac spinor $\psi$ may be written in the form as 
\be
\psi = C(r)^{-1} A(r)^{-1/2}~f(t, r)~\chi_{1} + C(r)^{-1} A(r)^{-1/2}~g(t, r)~\chi_{2}.
\label{dirsol}
\ee
Properties of {\it the hedgehog} spinors 
allow one \cite{gib93} to find that the {\it transverse} Dirac operator will act
as $\Dsl~\chi_{1} = - \chi_{2}$ and $\Dsl~ \chi_{2} = \chi_{1}$. Moreover 
one finds that $\bar{n} \bar{\sigma}\times \bar{\tau}~\chi_{1} =
- 2i \chi_{2}$ and $ \bar{n} \bar{\sigma}\times \bar{\tau}~ \chi_{2} = 2i \chi_{1}$. 
The above properties of {\it the hedgehog} spinors 
help us to rewrite 
the Dirac equation for the s-wave sector in the form
\ben \label{aa1}
\p_{t} f + \p_{r_{*}} g + {A(r)~( r - a(r)~C(r)) \over  r~ C(r)} g = 0, \\
\p_{t} g + \p_{r_{*}} f - {A(r)~( r - a(r)~C(r)) \over  r~ C(r)} f = 0,
\label{aa2}
\een
where for a convenience we have introduced the
{\it tortoise} coordinate connected with the $r$-coordinate by the relation $dr_{*}/dr = B(r)/A(r)$. 
By virtue of Eqs.(\ref{aa1})-(\ref{aa2}) 
we can readily conclude that they admit
a zero energy bound state given by
\be
f = exp \bigg ( \int_{r_{0}}^{r} {B(r)~( r - a(r)~C(r)) \over  r~ C(r)} dr \bigg),  \qquad g = 0.
\label{ff}
\ee
It will be interesting to find the near-horizon behaviour of the Dirac fermion fields. In the case
of a nonextremal black hole one can expand the metric
coefficients $A(r)$ and $B(r)$ in the vicinity of the black hole event horizon.
They will be provided by the following expressions:
\be
A(r)^2 \simeq A'(r_{+}) (r - r_{+}), \qquad
B(r)^2 \simeq B'(r_{+}) (r - r_{+})^{-1}.
\ee
Then, making a change of variables given by the relations
\be
\rho^2 = 4 B'(r_{+}) (r - r_{+}), \qquad T = {1 \over 2} \sqrt{{A'(r_{+}) \over B'(r_{+})}}~ t,
\ee
it can be shown that the line element describing the near horizon geometry of the nonextremal black hole
can be cast in the form as
\be
ds^2 = - \rho^2 dT^2 + d \rho^2 + C(r_{+})^2 (d\theta^2 + \sin^2 \theta d\phi^2 ).
\ee
Thus, one can approximate the spacetime 
in the vicinity of the nonextremal black hole event horizon by the Rindler line element.
On this account, having in mind Eqs.(\ref{dirsol}) and (\ref{ff}),
one can verify that the spinor function can be approximated by the following expression:
\be
\psi \simeq {1 \over C(r_{+})~A'(r_{+})^{1 \over 4}~(r - r_{+})^{1 \over 4}}.
\ee
One concludes that it behaves as $(r - r_{+})^{-1/4}$
near the event horizon $r_{+}$. If we shall consider the global monopole black hole spacetime
then $C(r_{+}) = b r_{+}$, where $b$ is the global monopole parameter. Just, the bigger $b$ we have
the more divergent is the spinor wave function. In the case of a dilaton black hole, one can remark that the 
$\alpha$-coupling constant will also trigger  the divergence of the spinor function near the black hole event horizon.

\par
On the other hand, in the extreme black hole case, when the outer black hole event horizon
is equal to the inner one, $r_{+} = r_{-}$, 
the metric functions imply the following:
\be
A(r)^2 \simeq {1 \over 2} A''(r_{+}) (r - r_{+})^2, \qquad
B(r)^2 \simeq {1 \over 2} B''(r_{+}) (r - r_{+})^{- 2}.
\ee
Let us use a suitable change of the coordinates given by
\be
\rho = \sqrt{{2 \over A''(r_{+})}}~{1 \over (r - r_{+})}, \qquad
T = {t \over \sqrt{2 B''(r_{+})}} = {t \over K}.
\ee
It can be verified that
in the coordinates $(T,~\rho,~\theta,~\phi)$ the line element of the near-horizon
extremal black hole metric becomes
\be
ds^2 = {K^2 \over \rho^2} \bigg( - dT^2 + d \rho^2 \bigg) + C(r_{+})^2 (d\theta^2 + \sin^2 \theta d\phi^2 ),
\ee
which is a Bertotti- Robinson type of the spacetime.
Making use of the near-horizon approximation 
we obtain the following value of the spinor function:
\be
\psi \simeq {1 \over {1 \over 2} A''(r_{+})^{1 \over 4}~ C(r_{+})~(r - r_{+})^{ \gamma - {1 \over 4}}},
\ee
where $\gamma$ implies
\be
\gamma = {({1 \over 2}~B''(r_{+}))^{1/4} \over C(r_{+})}.
\ee
Now, the spinor function diverges near the black hole event horizon as
$(r - r_{+})^{\ga - 1/4}$, where $\ga$ depends on the metric coefficients taken at $r_{+}$.
In the case under consideration, 
one has the influence of a global monopole parameter $b$ on the divergence
of the zero mode spinor function. In the case of the extremal dilaton black hole the situation is
much more complicated. As can be seen, for the extremal dilaton black hole 
the event horizon of it is singular in {\it Einstein frame} and has vanishing area.
Just $C(r_{+}) \rightarrow 0$ in this limit. However, in {\it string frame}
previously singular horizon has been pushed off to an infinite proper distance. The very
similar situation was indicated in the case of the expulsion of the Higgs vortex from an
extremal dilaton black hole \cite{mod99}. In the case under consideration
a full numerical study would be required.
To sum up, the divergence of spinor function for zero modes occurs in the near horizon limit
as was expected from the point of view of the {\it no-hair} theorem.
\par
Returning to relations (\ref{aa1})-(\ref{aa2}), we observe that they can be rewritten 
in the form of a coupled second order 
differential equations system.
Consequently, one obtains
\ben
\p_{t}^2 f - \p_{r_{*}}^2 f + H_{1}(r) f = 0, \\
\p_{t}^2 g - \p_{r_{*}}^2 g + H_{2}(r) g = 0, 
\een
where we have denoted by $H_{1}(r) = H^{2}(r) + \p_{r_{*}} H(r)$ and $H_{2}(r) = H^{2}(r) - \p_{r_{*}} H(r)$ 
the {\it effective} potentials for the s-wave sector
while {\it the potential} $H(r) = {A(r) (r - a(r) C(r)) \over r C(r)}$.
\par
One can remark that $H_{1}(r)$ and $H_{2}(r)$ are {\it supersymmetric} partners, in the sense 
presented in Ref.\cite{coo95},
derived from the same {\it superpotential} $H(r)$. In Ref.\cite{and91} it was proved that these relations between potentials 
provided that they were the sources of the same spectra of quasinormal modes. It means physically that Dirac fermions 
and antifermons have the same quasinormal modes in the considered 
general spherically symmetric black hole background.

\subsection{$k>0$ modes for Dirac fermions}
It happened
that the eigenspaces with eigenvalues $k$ and $m$ are four-dimensional ones for $k \geq 1$ \cite{gib93}. Operators appearing
in the Dirac equation can be expressed as matrices in the basis 
$( f_{L}|k,m,j_{+}>,~g_{L}|k,m,->,~ f_{R}|k,m,j_{-}>,~g_{R}|k,m,+>)$ in which the operators in question are diagonalized.
On this account, it can be readily seen that
\be
\psi = \pmatrix{\psi_{L} \cr \psi_{R}} =
C(r)^{-1} A(r)^{-1/2} 
\pmatrix{
f_{L}|k,m,j_{+}> \cr
g_{L}|k,m,-> \cr
f_{R}|k,m,j_{-}> \cr
g_{R}|k,m,+>
},
\ee
while the {\it transverse} Dirac operator implies the relation
\be
\Dsl = 
\pmatrix{
0 & 0 & 0 & k+1 \cr
0 & 0 & -k-1 & 0\cr
0 & -k & 0 & 0 \cr
k & 0 & 0 & 0}.
\ee
Also, we may note that the matrix form of the operator $\bar{n} \bar{\sigma}\times \bar{\tau}$ is 
given by the following:
\be
\bar{n} \bar{\sigma} \times \bar{\tau}=
{2i \over 2k+1}
\pmatrix{
-\sqrt{k(k+1)} & 0 & 0 & k+1 \cr
0 & \sqrt{k(k+1)} & -k-1 & 0 \cr
0 & - k & \sqrt{k(k+1)} & 0 \cr
k & 0 & 0 & -\sqrt{k(k+1)}}.
\ee
It can be verified that in this situation the Dirac equations reduce to the form
\ben
\p_{t} f_{R} &+& \p_{r^{*}} g_{R} + \beta(r) g_{R}  + \alpha(r) f_{L} = 0,\\
\p_{t} g_{R} &+& \p_{r^{*}} f_{R} - \beta(r) f_{R}  - \alpha(r) g_{L} = 0,\\
\p_{t} f_{L} &-& \p_{r^{*}} g_{L} - \beta(r) g_{L}  - \alpha(r) f_{R} = 0,\\
\p_{t} g_{L} &-& \p_{r^{*}} f_{L} + \beta(r) f_{L}  + \alpha(r) g_{R} = 0,
\een
where by $\alpha(r)$ and $\beta(r)$ we have denoted the following quantities:
\ben \label{aa}
\alpha(r) &=& {a(r)~ A(r) \over r (2k+1)} \sqrt{k(k+1)}, \\
\beta(r) &=& {A(r) \over C(r)}~D_{k} - {a(r)~ A(r) \over r(2k+1)}~D_{k}.
\label{bb}
\een
In Eq.(\ref{bb}) $D_{k}$ is equal to $k + 1$ for the $R$-function and equals to $k$ for the $L$-functions.
In order to simplify the above equations we put for all the functions an explicit time-dependence in the form $exp(-i\omega t)$.
We assume further, that the following relations between considered functions $f_{L} = if_{R}$ and 
$g_{L} = - ig_{R}$ are fulfilled.
It provides the following:
\ben \label{ww}
{d \over d r^{*}} \pmatrix{ g \cr f }
- {A(r) \over r} \pmatrix{
-\tilde{\beta}(r) & i \tilde{\alpha} \cr i \tilde{\alpha} & \tilde{\beta}(r)}
\pmatrix{ g \cr f } = 0,
 \een
where $\tilde{\alpha}(r)$ and $\tilde{\beta}(r)$ are given by
\ben
\tilde{\alpha}(r) &=& { \omega r \over A(r)} - {a(r)~ \sqrt{k(k+1)} \over 2k+1}, \\
\tilde{\beta}(r) &=& {r~(2k+1) - a(r)~ C(r) \over (2k+1)~C(r)}~(k+1).
\een
In order to simplify further the radial Eq.(\ref{ww}) let us make a change of variables defined as
\be
\pmatrix{ 
\tilde{g} \cr \tilde{f}} = T
\pmatrix{
g \cr f } =
\pmatrix{
\sin{\theta \over 2} & \cos{\theta \over 2} \cr
\cos{\theta \over 2} & -\sin{\theta \over 2}} 
\pmatrix{ g \cr f},
\ee
where the angle $\theta$ yields the following:
\be
\theta = arctan \bigg( {i \tilde{\alpha} \over \tilde{\beta}} \bigg).
\ee
Hence, the underlying relations reduce to the form as
\be
{d \over d r_{*}} \pmatrix{ \tilde{g} \cr \tilde{f}} - {A(r) \over r}~\sqrt{ \tilde{\beta}^2 - \tilde{\alpha}^2}
\pmatrix{1 & 0 \cr 0 & -1} \pmatrix{ \tilde{g} \cr \tilde{f}} =
- {1 \over 2 \omega} {d \theta \over d r_{*}}
\pmatrix{ 0 & - \omega \cr  \omega & 0} \pmatrix{ \tilde{g} \cr \tilde{f}},
\label{kkk}
\ee
It happens that they
may be simplified further. Namely, let us make another change of variables given by
$d \tilde {r} = {1 \over 2 \omega} {d \theta \over dr_{*}}$. Consequently, we arrive at
\be
{d \over d \tilde {r}} \pmatrix{ \tilde{g} \cr \tilde{f}} 
- W(r) \pmatrix{1 & 0 \cr 0 & -1} \pmatrix{ \tilde{g} \cr \tilde{f}} =
\pmatrix{ 0 & - \omega \cr  \omega & 0} \pmatrix{ \tilde{g} \cr \tilde{f}},
\ee
where by $W(r)$ we have denoted the following expression:
\be 
W(r) = {2 \omega A(r) \over r~ {d \theta \over dr_{*}}}~\sqrt{ \tilde{\beta}^2 - \tilde{\alpha}^2}.
\ee
The set of Eqs.(\ref{kkk}) can be decoupled providing the 
the system of second order differential equations. Namely, they yield
\ben
{d^2 \over d {\tilde {r}}^2} \tilde{g} &-& W_{1}(r)~ \tilde {g} = \omega^2~\tilde{g}, \\
{d^2 \over d {\tilde{r}}^2} \tilde{f} &-& W_{2}(r)~\tilde {f} = - \omega^2~\tilde{f},
\een
where the {\it effective} potentials $W_{1}(r)$ and $W_{2}(r)$ imply
\ben
W_{1}(r) = W^{2}(r) + \p_{\tilde {r}} W(r), \\
W_{2}(r) = W^{2}(r) - \p_{\tilde {r}} W(r).
\een
Because of the fact that $W_{1}(r)$ and $W_{2}(r)$ are {\it supersymmetric} to each other,
$\tilde{g}$ and $\tilde{f}$ will have the same spectra both for quasinormal modes and
scattering.

\subsection{Massive Dirac Fermion Modes}
So far we have considered massless fermion case. Now, we wish to generalize our considerations 
and present some relevant arguments concerning with
a massive Dirac fermion.
The equation under consideration implies
\be
i~\gamma^{\mu}(\nabla_{\mu}-i\lambda H_{\mu})\psi -m\psi=0,
\label{mass}
\ee
where $m$ is a mass of the Dirac fermion. By the same procedure that we followed in the preceding section,
the Dirac equations for the massive case can be written as
\ben
\p_{t} f_{R} &+& \p_{r^{*}} g_{R} + \beta(r) g_{R}  + \alpha(r) f_{L} + i~m~A(r) f_{L}= 0,\\
\p_{t} g_{R} &+& \p_{r^{*}} f_{R} - \beta(r) f_{R}  - \alpha(r) g_{L} + i~m~A(r) g_{L}= 0,\\
\p_{t} f_{L} &-& \p_{r^{*}} g_{L} - \beta(r) g_{L}  - \alpha(r) f_{R} + i~m~A(r) f_{R}= 0,\\
\p_{t} g_{L} &-& \p_{r^{*}} f_{L} + \beta(r) f_{L}  + \alpha(r) g_{R} + i~m~A(r) g_{R}= 0,
\een
Further, assuming that $f_{L} =  i f_{R}$ and $g_{L} = - i g_{R}$, we obtain
\be
{d \over d r^{*}} \pmatrix{f \cr g}
-
\pmatrix{
 \beta(r) & \tilde{\alpha}_{1} \cr \tilde{\alpha}_{1} & - \beta(r)}
\pmatrix{f \cr g}
=
\pmatrix{
0 & -m~A(r) \cr m~A(r) & 0}
\pmatrix{ f \cr g},
\ee
where $\tilde{\alpha}_{1}(r)$ is provided by
\be
\tilde{\alpha}_{1} = i \bigg( \omega - {a(r)~A(r)~ \sqrt{k(k+1)} \over r (2k+1)}
\bigg).
\ee
As in the previous section, let us introduce 
$\theta(r)$ defined as
\be
\theta(r) = arctan \bigg( {\tilde{\alpha}_{1} \over \beta(r) } \bigg).
\ee
Applying the transformation $T$ and introducing new variable defined as
$d {\tilde r} = \bigg( A(r) - (1/ 2m) { d \theta \over dr_{*}} \bigg)~dr_{*}$, 
we arrive at the following expression:
\be
{d \over d {\tilde r}}
\pmatrix{ \tilde{f} \cr \tilde{g}} 
-
{ \sqrt{ \beta(r)^2 + \tilde{ \alpha}_{1}(r)^2}  \over 
\bigg( A(r) - {1 \over  2m}~ {d \theta \over dr_{*}} \bigg)}
\pmatrix{ \tilde{f} \cr - \tilde{g}}
=
- m 
\pmatrix{ \tilde{g} \cr - \tilde{f}}. 
\ee
It can be also shown that these equations decouple. Namely, one gets
\ben
\p_{\tilde r}^2~{\tilde f} &-& G_{1}(r)~{\tilde f} + m^2~{\tilde f} = 0, \\
\p_{\tilde r}^2~{\tilde g} &-& G_{2}(r)~{\tilde g} + m^2~{\tilde g} = 0, 
\een
where $G_{1}(r)$ and $G_{2}(r)$ yield
\be
G_{1}(r) = \p_{\tilde r}~W + W^2, \qquad G_{2}(r) = - \p_{\tilde r}~W + W^2,
\ee
while by $W(r)$ we have denoted
\be
W = {\sqrt{\beta (r)^2 + {\tilde \alpha}_{1} (r)^2} \over A(r) -
{1 \over  2m}~ {d \theta \over dr_{*}}}.
\ee
The above potentials have the same features as potentials 
in preceding sections. Namely, they are {\it supersymmetric}
to each other.

\section{The Decay of Dirac Fermion Hair }

This section will be devoted to the problem
of the decay of massive Dirac fermion hair
in the background of black holes with non-trivial topology of the event horizon.
We shall use the sign convention presented in Ref. \cite{gibrog08}, where it was pointed out that
that the
treatment of fermions in spherically symmetric backgrounds may be
greatly simplified by recalling a few basic properties of the Dirac
equation.
Namely, for a line element of the form as
$g_{\mu \nu} d x ^\mu d x ^\nu = 
 g_{ab} (x) dx^a dx ^b + g_{mn}(y) dy^m dy ^n$ the Dirac operator
$\Dsl = \gamma^\mu \nabla _\mu $ can be decomposed
as a direct sum 
\be
\Dsl = \Dsl_x + \Dsl_y.
\ee
Moreover, under a Weyl conformal rescaling of metric tensor  
$
g_{\mu \nu} = \Omega ^2 {\tilde g} _{\mu \nu}$
it could be shown that
\be
\Dsl \psi =\Omega ^{- {1 \over 2} (n+1)} { \tilde {\Dsl} } {\tilde \psi}\,,  
\qquad \psi = \Omega ^ {- {1 \over 2}  (n-1) }\tilde \psi .
\ee
Let us consider a conformo-static metric of the form as
\be
ds^2 =-A^2 dt^2 +\Phi ^2 d x ^i d x^i\,,
\ee
where $A=A(x^i)$ and $\Phi=\Phi(x^i)$, $i=1,2\dots n-1$. Then, one obtains 
\be
ds^2 =A^2 \biggl( - dt^2 +\bigl ( {\Phi \over A} \bigr) ^2 d x ^i d x^i 
\biggr)\,,
\ee
and finally we may note that
\be
\Dsl \psi = A^{-{1 \over 2} (n+2)}  \biggl
( \gamma ^0 \partial_t  +\tilde \Dsl  \biggr) \tilde \psi,
\ee
where $\tilde \Dsl$ is the Dirac operator of the metric
$ \bigl ( {\Phi \over A} \bigr) ^2 d x ^i d x^j$ and 
$\tilde \psi= A^{{1 \over 2} (n-1) } \psi$.
Using again the conformal property one arrives at the following:
\be
\tilde \Dsl \tilde \psi = \bigl( {A  \over \Phi }\bigr) ^{ {1 \over 2} (n-1) }
 \gamma ^i \partial_i
\tilde{\tilde \psi}\,, 
\ee 
with $\tilde \psi =  \bigl( {A  \over \Phi } \bigr ) ^{{1\over 2}(n-2)} 
\tilde {\tilde \psi}$.
Let us suppose moreover, that $\Psi$ is a spinor eigenfunction on the 
$(n-2)$-dimensional {\it transverse} manifold $\Omega$ satisfying relation
\be
\Dsl_\Omega \Psi = \lambda \Psi.
\label{prop1}
\ee
In the case of a $(n - 2)$-dimensional sphere, the eigenvalues of the spinor $\Psi$
were given in Ref.\cite{cam96} in the form
\be
\la^2 = \bigg( l + { n - 2 \over 2} \bigg)^2,
\ee
where $l = 0, 1, \dots$
\par  
Furthermore, one can assume that $\Dsl \psi = m \psi $
and set the following form of $\psi$:
\be
\psi = {1 \over A^{1 \over 2} } { 1 \over C^{(n-2) \over 2 }} \chi \otimes \Psi.
\label{prop2}
\ee
It can be checked by the direct calculations that the above form of fermion fields provides
the following:
\be
(\gamma ^0 \partial _t + \gamma ^1 \p_{r_{*}}) \chi = 
A(r)~ \bigg( m- {\lambda \over C(r)} 
\bigg) \chi.  
\ee
One should recall that
the matrices $\gamma ^0, \gamma ^1$ 
satisfy the Clifford algebra in two spacetime dimensions.
If we assume
that $\psi \propto e^{-i\omega t}$ it can be shown that we are left with the
second order equation of the form
\be
{d^2 \chi \over d y^2 }+ \omega ^2 \chi  = 
A(r)^2 \bigg ( m-{\lambda \over C(r)} \bigg )^2 \chi. 
\label{second}
\ee
Although the above derivations are valid for arbitrary number of spacetime
dimensions, in what follows we shall restrict our attention to the four-dimensional case.

\subsection{The background of a black hole with a global monopole}
We first focus on the case of decaying massive fermion hair on the background of a black hole with a global monopole.
In four-dimensional spacetime the line element describing a black hole which swallowed a global monopole
is written as
\be
ds^2 = - \bigg( 1 - 8 \pi G \eta^2 - {2 G \tM \over r}
 \bigg) dt^2
+ {dr^2 \over \bigg( 1 - 8 \pi G \eta^2 - {2 G \tM \over r}
 \bigg)}
+ r^2 d\Omega^2,
\label{mon}
\ee
where by $\tM$ we denote mass of black hole and $\eta$ is the symmetry breaking scale when the monopole is produced.
If we introduce the coordinate transformation in the form
\be
t \rightarrow (1 - 8 \pi G \eta^2)^{-3/2} t, \qquad
r \rightarrow (1 - 8 \pi G \eta^2)^{-1/2} r,
\ee
as well as the new parameters which yield
\be
M = (1 - 8 \pi G \eta^2)^{-3/2} \tM, \qquad b^2 = 1 - 8 \pi G \eta^2,
\ee
then, the line element of a black hole with global monopole reduces to the form given by Eq.(\ref{monopol}).
In what follows we put $G = 1$ for simplicity.
\par
The spectral decomposition method will be our main tool
in the analysis of the time evolution of a massive Dirac spinor field in 
the background of a black hole with global monopole.
It was revealed in
Refs.\cite{lea86} that the asymptotic tail was connected with the
existence of a branch cut situated along the interval $-m \le \omega \le
m$. An oscillatory inverse power-law behaviour of a~massive Dirac field
arises from the integral of the Green function $\tilde G(y, y';\omega)$
around the branch cut. The time evolution of the massive Dirac field
takes the form
\be
\chi(y, t) = \int dy' \bigg[ G(y, y';t) \chi_{t}(y', 0) +
G_{t}(y, y';t) \chi(y', 0) \bigg],
\ee
for $t > 0$, where the Green's function  $ G(y, y';t)$ implies the following relation:
\be
\bigg[ {\p^2 \over \p t^2} - {\p^2 \over \p y^2 } + V \bigg]
G(y, y';t)
= \delta(t) \delta(y - y'),
\label{green}
\ee
where $V$ is an effective potential.\\
In order to find the Green function in the case under consideration we shall
use the Fourier transform \cite{lea86}
$\tilde  
G(y, y';\omega) = \int_{0^{-}}^{\infty} dt~ G(y, y';t) e^{i \omega t}$ and reduce
Eq.(\ref{green}) to an ordinary differential one.
The Fourier's transform is well defined for $Im~ \omega \ge 0$, while
the inverse transform becomes
\be
G(y, y';t) = {1 \over 2 \pi} \int_{- \infty + i \ep}^{\infty + i \ep}
d \omega~
\tilde G(y, y';\omega) e^{- i \omega t},
\ee
for some positive number $\ep$.
Thus, in this picture
the above Fourier's component of the Green function $\tilde  G(y, y';\omega)$
can be rewritten in terms of two linearly independent solutions of the
homogeneous equation of the form
\be
\bigg(
{d^2 \over dy^2} + \omega^2 - \tV \bigg) \chi_{i} = 0, \qquad i = 1, 2,
\label{wav}
\ee
where we have  denoted by $\tV = A(r)^{2} \bigg( m - {\la \over C(r)} \bigg)^2$.\\
Let us consider the boundary conditions of the problem in question.
For $\chi_{i}$ they are described by purely ingoing waves
crossing the outer horizon $H_{+}$ of the 
static charged black hole
$\chi_{1} \simeq e^{- i \omega y}$ as $y \rightarrow  - \infty$. 
As far as 
$\chi_{2}$ is concerned, it
should be damped exponentially at $i_{+}$, i.e.,
$\chi_{2} \simeq e^{- \sqrt{m^2 - \omega^2}y}$ at $y \rightarrow \infty$.
\par
In our considerations we shall assume that
the observer and the initial data are situated far away from the black hole
with global monopole. 
Eq.(\ref{wav}) may be rewritten by using new variables expressed as
\be
\chi_{i} = {\xi \over \bigg( 1 - {2 M \over r} \bigg)^{1/2}},
\ee
where $i = 1,2$. 
Next, we expand
Eq.(\ref{wav}) in a power  series of $ 2 M /r$ neglecting terms of order
$\cO ((\omega/r)^2)$ and higher. Having this in mind we are left with the equation
\be
{d^2 \over dr^2} \xi + \bigg[
\omega^2 - m^2 + {4 M b \omega^2  + 2 \la m \over b r} -{\la^2 \over b^2 r^2}
\bigg] \xi = 0.
\label{whit1}
\ee
The main result of the above procedure is the conclusion that 
Eq.(\ref{whit1}) may be solved in terms of Whittaker's functions. 
Two basic solutions are needed to construct the Green function, with the condition that
$\mid \omega \mid \ge m$. 
The Whittaker's functions provided the solution of the above equation are
$\tchi_{1} = M_{\delta, \tim}(2 \tom r)$ and $\tchi_{2} = W_{\delta, \tim}(2 \tom r)$,
while their parameters are given by
\be
\tim = \sqrt{ 1/4 + {\la^2 \over b^2}}, \qquad    \delta =  {4 M b \omega^2  + 2 \la m \over 2 \tom b},
 \qquad        
\tom^2 = m^2 - \omega^2.
\ee
It is sufficient to conclude that the spectral Green function takes the form as
\ben
G_{c}(r, r';t) &=& {1 \over 2 \pi} \int_{-m}^{m}dw
\bigg[ {\tchi_{1}(r, \tom e^{\pi i})~\tchi_{2}(r',\tom e^{\pi i}) \over W(\tom e^{\pi i})}
- {\tchi_{1}(r, \tom )~\tchi_{2}(r',\tom ) \over W(\tom )} 
\bigg] ~e^{-i w t} \\ \nonumber
&=& {1 \over 2 \pi} \int_{-m}^{m} dw f(\tom)~e^{-i w t},
\een 
where $W(\tom)$ is the Wronskian.\\
We first focus our attention on
the intermediate asymptotic decay of the massive Dirac hair, i.e., in the range of parameters
$M \ll  r \ll t \ll M/(m M)^2$.
The intermediate asymptotic contribution to the spectral Green function integral gives the frequency equal to 
$\tom = {\cO (\sqrt{m/t})}$, which in turns implies that $\delta \ll 1$. Having in mind that $\delta$ 
results from the $1/r$ term in the massive Dirac field equation of motion, it depicts
the effect of backscattering off the spacetime curvature and in the case under consideration
the backscattering is negligible. Taking into account all the above we obtain the result
\be
f(\tom) = {2^{2 \tim -1} \Gamma(-2\tim)~\Gamma({1 \over 2} + \tim) \over
\tim \Gamma(2 \tim)~\Gamma({1 \over 2} - \tim)} \bigg[
1 + e^{(2 \tim + 1) \pi i} \bigg]
(r r')^{{1 \over 2} + \tim} \tom^{2 \tim},
\ee
where we have used the fact that $\tom r \ll 1$ and the form of $f(\tom)$
can be approximated by means of the fact that $M(a, b, z) = 1$ as $z$ tends to zero.
The resulting Green function reduces to the form as
\be
G_{c}(r, r';t) = {2^{3 \tim - {3\over2}} \over \tim \sqrt{\pi}}
{\Gamma(-2\tim)~\Gamma({1 \over 2} + \tim) \Gamma(\tim +1 ) \over
\tim \Gamma(2 \tim)~\Gamma({1 \over 2} - \tim)}
\bigg( 1 + e^{(2 \tim + 1) \pi i} \bigg)~(r r')^{{1 \over 2} + \tim} 
~\bigg( {m\over t} \bigg)^{{1 \over 2} + \tim}~J_{{1 \over 2} + \tim}(mt).
\ee  
In the limit when $t \gg 1/m$ one can show that the spectral Green function yields
\be
G_{c}(r, r';t) = {2^{3 \tim - 1} \over \tim \sqrt{\pi}}
{\Gamma(-2\tim)~\Gamma({1 \over 2} + \tim) \Gamma(\tim +1 ) \over
\tim \Gamma(2 \tim)~\Gamma({1 \over 2} - \tim)}
\bigg( 1 + e^{(2 \tim + 1) \pi i} \bigg)~(r r')^{{1 \over 2} + \tim} 
~m^{\tim}~ t^{- 1 - \tim}~\cos(mt - {\pi \over 2}(\tim + 1)).
\label{gfim}
\ee  
Eq.(\ref{gfim}) depicts the oscillatory inverse power-law behaviour. In our case the intermediate
times of the power-law tail depend only on $\tim$ which in turn is a function of the multipole number
of the wave modes and monopole parameter $b$.
\par
On the other hand, for the late-time behaviour quite different pattern of the decay should be expected because of the fact
that backscattering off the curvature of the spacetime play an important role.
For this case $\kappa \gg 1$ and $f(\tom)$ may be rewritten using the fact that
$M_{\delta, \tim}(2 \tom r) \approx
\Gamma (1 + 2 \tim) (2 \tom r)^{1 \over 2}~\delta^{- \tim}~J_{\tim}(\sqrt{8 \delta \tom r})$.
It yields
\ben \label{fer}
f(\tom) &=& {\Gamma(1 + 2\tim)~\Gamma(1 - 2\tim) \over 2 \tim}~(r r')^{1 \over 2}
\bigg[ J_{2 \tim} (\sqrt{8 \delta \tom r})~J_{- 2 \tim} (\sqrt{8 \delta \tom r'})
- I_{2 \tim} (\sqrt{8 \delta \tom r})~I_{- 2 \tim} (\sqrt{8 \delta \tom r'}) \bigg] \\ \nonumber
&+&
{(\Gamma(1 + 2\tim))^2~\Gamma(- 2\tim)~\Gamma( {1 \over 2} + \tim - \delta)
 \over 2 \tim ~\Gamma(2 \tim)~\Gamma({1 \over 2} - \tim - \delta) }~(r r')^{1 \over 2}
~\delta^{- 2 \tim}
\bigg[
J_{2 \tim} (\sqrt{8 \delta \tom r})~J_{2 \tim} (\sqrt{8 \delta \tom r'})
\\ \nonumber
&+& e^{(2 \tim + 1)}
I_{2 \tim} (\sqrt{8 \delta \tom r})~I_{2 \tim} (\sqrt{8 \delta \tom r'}) 
\bigg].
\een
The first part of the above Eq.(\ref{fer}) the late time tail is proportional to $t^{-1}$
and it occurs that we shall concentrate on
the second term of the right-hand side of Eq.(\ref{fer}). For the case when 
$\kappa \gg 1$ it can be brought to the standard form
written as
\be
G_{c~(2)}(r, r';t) = {N \over 2 \pi} \int_{-m}^{m}~dw~e^{i (2 \pi \delta - wt)}~e^{i \varphi},
\ee
where we have defined
\be
e^{i \varphi} = { 1 + (-1)^{2 \tim} e^{- 2 \pi i \delta} \over
 1 + (-1)^{2 \tim} e^{2 \pi i \delta}},
\ee
while $N$ provides the relation as follows:
\be
N = {(\Gamma(1 + 2\tim))^2~\Gamma(- 2\tim) \over 2 \tim ~\Gamma(2 \tim) }~(r r')^{1 \over 2}
\bigg[
J_{2 \tim} (\sqrt{8 \delta \tom r})~J_{2 \tim} (\sqrt{8 \delta \tom r'})
+ I_{2 \tim} (\sqrt{8 \delta \tom r})~I_{2 \tim} (\sqrt{8 \delta \tom r'}) 
\bigg].
\ee
At very late time both terms $e^{i w t}$ and $e^{2 \pi \delta}$ are rapidly
oscillating. It means that the scalar waves are mixed states consisting of the states 
with multipole phases backscattered by spacetime curvature. Most of them cancel
with each others which have the inverse phase. In such a case, one can find the value of 
$G_{c (2)}$ by means of the saddle point method. It can be found that
the saddle point is given by
\be
a_{0} = \bigg[ {\pi~(4 M b \omega^2~ + 2 \la m) \over  2 \sqrt{2} b m} \bigg]^{1 \over 3},
\ee
In comparison to the late-time behaviour of the second term in Eq.(\ref{fer}), the first term 
can be neglected. The dominant role plays the behaviour of the second term, i.e., the late-time 
behaviour is proportional to
${- {5 \over 6}}$. 
Consequently, this fact implies the resultant form of the spectral Green function for the late-time
behavior of massive Dirac field in black hole with global monopole spacetime, namely it can be written as
\be            
G_{c}(r, r';t) = { 2 \sqrt{2}  b^{1 \over 3} \over \sqrt{3}}~(\pi)^{5 \over 6}
\bigg[
4 M b m^2 + 2 \la m
\bigg]^{1 \over 3}
(mt)^{-{ 5 \over 6}}~\sin(mt)~\tchi(r, m)~\tchi(r', m),
\ee
One can see that the late-time behaviour is independent of the global monopole parameters
as well as the mass of the Dirac field.

\subsection{The background of a black hole with a cosmic string}
In this subsection we analyze the decay pattern of massive Dirac hair on the spherically symmetric solution 
of dilaton gravity being the low-energy limit of the string theory.
In four spacetime dimensions,  the action for the dilaton gravity 
with arbitrary coupling constant $\alpha$ is given by
\be
S = \int d^4x \sqrt{-g} 
\bigg[ R - 2 \na^{\mu} \phi \na_{\mu} \phi - e^{- 2 \alpha \phi} 
F_{\mu \nu}  F^{\mu \nu} \bigg],
\ee
where $\phi$ is the dilaton field, 
$F_{\mu \nu} = 2 \na_{[\mu} A_{\nu]}$
is the strength of the $U(1)$ gauge field.\\
This theory constitutes the low-energy approximation of the heterotic string theory.
The static spherically symmetric black hole solution with a cosmic string passing through is a 
{\it thin string} approximation of the Nielsen-Olesen vortex piercing the black hole.
The metric for such a system was found in
Ref.\cite{mod99}. It was assumed that the system underlies the complete separation of the degrees of freedom
between the $U(1)$ gauge field responsible for the charge of the black hole in question and the gauge vortex field. 
\par
The line element we shall consider will be given by Eq.(\ref{dila}).
The same procedure as in the preceding section and Ref.\cite{gibrog08}
enables us to find the intermediate and late-time asymptotic behaviour of massive Dirac hair
in the background of dilaton black hole with a cosmic string. For the completeness
and the reader's convenience we describe the crucial points of the underlying procedure.
Namely, it is
convenient to change variables
in Eq.(\ref{wav}) in the way as follows:
\be
\chi_{i} = {\xi \over \bigg( 1 - {r_{+} \over r} \bigg)^{1 \over 2}
\bigg( 1 - {r_{-} \over r} \bigg)^{{1 - \alpha^2} \over 2(1 + \alpha^2)}},
\ee
where $i = 1,2$. 
Neglecting the terms of order $\cO ((\omega/r)^2)$ and higher leads us to the relation for $\xi$
\ben \label{whit}
{d^2 \over dr^2} \xi &+& \bigg[
\omega^2 - m^2 + {2 \omega^2 ( r_{+} + \alpha_{1} r_{-}) 
- m^2 ( r_{+} + \alpha_{1} r_{-})
+ 2 \la m (1 + r_{+})
\over r} \\ \nonumber
&-& {\la^2 - 2 \la m r_{-} ( \alpha_{1} + \alpha_{2}) + m^2 \alpha_{1} r_{+} r_{-}
\over r^2}
\bigg] \xi = 0,
\een
which in turns provides the solutions given by Whittaker's functions. Namely,
$\tchi_{1} = M_{\delta, \tim}(2 \tom r)$ and $\tchi_{2} = W_{\delta, \tim}(2 \tom r)$ with the
following parameters:
\ben \label{par1}
\tim &=& \sqrt{ 1/4 + \la^2 - 2 \la m r_{-} + m^2 \alpha_{1} r_{+} r_{-} }, \\ \nonumber
\delta &=&  {\omega^2 ( r_{+} + \alpha_{1} r_{-}) + \la m (1 + r_{+}) - {m^2 \over 2}( r_{+} + \alpha_{1} r_{-})
\over \tom},\\ \nonumber
\tom^2 &=& m^2 - \omega^2.
\een
The intermediate asymptotic behaviour of the massive Dirac hair on the dilaton black hole pierced by a cosmic 
string will be given by Eq.(\ref{gfim}), but in this case the parameters 
of Whittaker's functions
$\tim$ and $\delta$ are of the 
form provided by the relation (\ref{par1}).
\par
As far as the late-time asymptotic behaviour is concerned it yields
\be            
G_{c}(r, r';t) = { 2 \sqrt{2} \over \sqrt{3}}~m^{2/3}~ (\pi)^{5 \over 6}
\bigg[
2 m^2 (r_{+} + \alpha_{1} r_{-}) + 2 \la m (1 + r_{+}) - m^2 (r_{+} + \alpha_{1} r_{-}) 
\bigg]^{1 \over 3}
(mt)^{-{ 5 \over 6}}~\sin(mt)~\tchi(r, m)~\tchi(r', m).
\label{spg}
\ee
It envisages the fact that
the late-time asymptotic decay pattern of massive Dirac hair in the background of spherically symmetric
dilaton black hole is proportional to $- 5/6$.
\par
Having in mind the properties of the Dirac operator presented in the preceding sections, we can
see from the relations (\ref{prop1})-(\ref{prop2}) that the main modification will stem from
the quite different topology of the {\it transverse} manifold. In our case it will be $S^2$-sphere
with a deficit angle. The reason of it is the presence of
a cosmic string passing through it. The other part of the Dirac function will be not
affected by the string. 
Now we proceed to the analysis of the eigenvalues of the Dirac operator on the underlying manifold.\\
The metric on the sphere with cosmic string has the form as:
\be
d \tOm^2 = B^2 \sin^2 \theta d^2 \phi + d^2 \theta,
\ee
where $B = 1 - 4 \mu$, while $\mu$  is the mass per unit length of the cosmic string \cite{ary86}.\\
As can be seen by the direct calculations the Dirac operator may be written as follows:
\be
\Dsl_{\tS^2} \psi = \ga^{1} \p_{\theta} + {\ga^{2} \over B \sin \theta} \p_{\phi} \psi
+ {\ga^2 \cot \theta \over 4}[\ga^{2}, \ga^{1}]~\psi,
\label{dop}
\ee
where gamma matrices satisfy the Clifford algebra rules in two-spacetime dimensions.
\par
The square of ${\Dsl}^2_{\tS^2}$ takes the form
\be
{\Dsl}^2_{\tS^2} = {\p_{\theta} (\sin \theta~\p_{\theta}) \over \sin \theta}
+ {1 \over B^2 \sin^2 \theta} {\p_{\phi}}^2 -
{\cos \theta \over B \sin^2 \theta}~i ~\sigma_{3}~\p_{\phi}
- {1 \over \sin^2 \theta} - {1 \over 4}.
\ee
Let us assume that
the eigenfunctions of the Dirac operator (\ref{dop}) are two-component spinors that fulfilled
the following:
\be
\pmatrix{
\alpha_{\la}(\theta, \phi) 
\cr \beta_{\la}(\theta, \phi) \cr}
= \sum_{m} {e^{i m \phi} \over \sqrt{2}}~
\pmatrix{
\alpha_{\la m}(\theta) 
\cr \beta_{\la m}(\theta) \cr}
\ee
where $m$ are half-integers.
On evaluating
the Dirac equation on a {\it stringy sphere} we find that it provides the system of differential
equations expressed as
\ben
\bigg( \p_{\theta} + {\cot \theta \over 2} \bigg) \beta_{\la m} ( \theta)
&+& {\tm \over \sin \theta} \beta_{\la m} ( \theta) = \la \alpha_{\la m} ( \theta), \\ \noindent
\bigg( \p_{\theta} + {\cot \theta \over 2} \bigg) \alpha_{\la m} ( \theta)
&-& {\tm \over \sin \theta} \alpha_{\la m} ( \theta) = \la \beta_{\la m} ( \theta),
\een
where $\tm = m / B$.
Next, we change the variables $x = \cos \theta$. It allows us to rewrite the underlying relations as
\be
\bigg[
{d \over dx}\bigg( 1 - x^2 \bigg) - {\tm^2 - \tm~ \sigma_{3}~x + {1 \over 4} \over 1 - x^2}
\bigg]~\pmatrix{
\alpha_{\la m}(\theta) 
\cr \beta_{\la m}(\theta) \cr} 
= - \bigg( \la^2 - {1 \over 4} \bigg)~\pmatrix{
\alpha_{\la m}(\theta) 
\cr \beta_{\la m}(\theta) \cr},
\ee
where $\sigma_{3}$ is the Pauli matrix.
The above equations are singular at the poles of the sphere $x = \pm 1$, so 
it is convenient to redefine the unknowns
\be
\pmatrix{
\alpha_{\la m}(\theta) 
\cr \beta_{\la m}(\theta) \cr} =
\pmatrix{
(1 - x)^{{1 \over 2} \mid \tm - {1 \over 2} \mid} ~(1 + x)^{{1 \over 2} \mid \tm + {1 \over 2} \mid}~\xi_{\la m}(x)
\cr
(1 - x)^{{1 \over 2} \mid \tm + {1 \over 2} \mid} ~(1 + x)^{{1 \over 2} \mid \tm - {1 \over 2} \mid}~\eta_{\la m}(x)
\cr}.
\ee
It can be verified by evaluating these expressions, that one can achieve to the relations
\be
\bigg[
\bigg( 1 - x^2 \bigg)~{d^2 \over dx^2}
+ \bigg(
{\tm \over \mid \tm \mid}~\sigma_{3} - \bigg( 2 \mid \tm \mid + 2 \bigg)~x
\bigg) {d \over dx} - \tm (\tm + 1) + \bigg( \la^2 - {1 \over 4} \bigg)
\bigg]~\pmatrix{
\xi_{\la m}(x)
\cr \eta_{\la m}(x)  
 \cr} = 0.
\ee
Having in mind the general formula for Jacobi polynomials (see e.g., \cite{abr70}) one gets 
the square of the eigenvalues for the Dirac operator on the $S^2$-sphere with a cosmic string passing through it,
which yields
\be
\la^2 = \bigg( l + \mid \tm \mid + {1 \over 2} \bigg)^2,
\ee
where $l = 0, 1, 2 \dots$
Because of the fact that $\tm = m / B$
the crucial role is played by the factor $B$ connected with a mass per unit length of a string.
\par
To conclude this section we remark that recently the non-linear origin of the power law tail in the long-time
evolution of a spherically symmetric self-gravitating massless scalar field was discussed (\cite{biz09} and references therein).
The perturbation method was used to obtain the expression for the tail 
and then numerical integration was performed to check the results.
The non-linear evolution of a black ringdown in the framework
of the higher-order metric perturbation theory was conducted in Ref.\cite{oku08}. 
It was argued that these non-linear components should be particularly significant for binary
black hole coalescences.
\par
Although, it is undoubtedly that the linear theory is clearly useful \cite{lea86,linth}, the recent
numerical simulations of the aforementioned problems yield that non-linearity is also worth elaborating.
We hope to return to the problem of non-linear behaviour of massive Dirac hair elsewhere.

\section{Conclusions}
To summarize, we have analyzed
the behaviour of Dirac fermions in the background of non-trivial topologies.
Assuming the complete separation of the degrees of freedom of the fields in question,
we
have first considered the massless zero modes in the near horizon limit for 
both nonextremal and extremal black holes. We have found that the spinor function is divergent near horizon as
$(r - r_{+})^{-1/4}$ for the non-extremal case. In the vicinity of extremal black hole horizon one gets the dependence
$(r - r_{+})^{\gamma -1/4}$, where $\gamma$ was built from adequate components of line element taken at
$r_{+}$. It was also established that the global monopole parameter influenced the divergent behaviour. 
Namely,
the bigger it is the more divergent the spinor wave function is.
On the other hand, in dilaton gravity black hole spacetime one has the coupling constant
influence on the divergence of the spinor wave function.
\par
Then, we take into account Dirac fermion modes for $k > 0$ and show that the Dirac 
equations can be decoupled
to the system of second order differential equations. We treat also the massive case of Dirac fermion fields.
As in the massless case one has also the situation that underlying equations decouple
to the system of second order differential equations with the so-called {\it supersymmetric}
potentials.
\par
Next, we proceed to the intermediate and late-time behaviour of massive Dirac hair
in the backgrounds of black holes with topological defects. Namely, we considered the black hole with
global monopole and cosmic string passing through it. In the case of black hole which swallowed a
global monopole one gets the modification of the intermediate late-time behaviour which depends 
on mass of the field in question as well as the global monopole mass. The intermediate late-time
decay of the hair is quicker comparing to the decay rate of massive Dirac hair on a black hole
without global monopole. The intermediate oscillatory power-law depends also on the multipole number of the wave mode.
But it is not the final pattern of the decay of the adequate massive hair. At very late times
the resonance backscattering off the spacetime curvature plays the dominant role. This decay pattern
is independent of the presence of a monopole and of the mass of the Dirac field, and it is
proportional to $t^{- 5/6}$.
\par
The analysis of the decay of massive Dirac hair on the background of a dilaton black hole with a cosmic
string passing through it, reveals that the intermediate as well as the late-time behaviour is
independent on the presence of this kind of topological defect. 
The main modification appears in the eigenvalues of the Dirac
operator on a $S^2$-sphere pierced by the cosmic string. The cosmic string parameter connected with its mass per unit 
length plays the crucial role in this case.
On the other hand, the intermediate late-time behaviour depends on the multipole number of the
wave mode as well as the mass of the Dirac field.
In turn, the late-time behaviour of the massive Dirac field is independent of the above factors and
it is proportional to $t^{- 5/6}$. 
\par
Having in mind the previous works treated the scalar, fermion and vector black hole hair decays
one can conclude that at asymptotically late-times the decay of the hair in question is universal
and does not depend on spin of the field, wave number of the mode and the topology of the black hole 
event horizon.



\begin{acknowledgments}
This work was partially financed by the budget funds in 2010 as research project.
\end{acknowledgments}


\begin{thebibliography}{99}
%
\def\cmp#1#2#3{{ Commun. Math. Phys.} {\bf #1}, #2 (#3)}
\def\lmp#1#2#3{{ Lett. Math. Phys.} {\bf #1}, #2 (#3)}
\def\hpa#1#2#3{{ Hell. Phys. Acta} {\bf #1}, #2 (#3)}
\def\grg#1#2#3{{ Gen. Rel. Grav.} {\bf #1}, #2 (#3)}
\def\pr#1#2#3{{ Phys. Rev.} {\bf #1}, #2 (#3)}
\def\prl#1#2#3{{ Phys. Rev. Lett.} {\bf #1}, #2 (#3)}
\def\prd#1#2#3{{ Phys. Rev. D} {\bf #1}, #2 (#3)}
\def\pl#1#2#3{{ Phys. Lett} {\bf #1}, #2 (#3)}
\def\pla#1#2#3{{ Phys. Lett. A} {\bf #1}, #2 (#3)}
\def\plb#1#2#3{{ Phys. Lett. B} {\bf #1}, #2 (#3)}
\def\prep#1#2#3{{ Phys. Reports} {\bf #1}, #2 (#3)}
\def\phys#1#2#3{{ Physica} {\bf #1}, #2 (#3)}
\def\jcp#1#2#3{{ J. Comput. Phys.} {\bf #1}, #2 (#3)}
\def\jmp#1#2#3{{ J. Math. Phys.} {\bf #1}, #2 (#3)}
\def\jpm#1#2#3{{ J. Phys. A: Math. Gen.} {\bf #1}, #2 (#3)}
\def\cpr#1#2#3{{ Computer Phys. Rept.} {\bf #1}, #2 (#3)}
\def\cqg#1#2#3{{ Class. Quantum Grav.} {\bf #1}, #2 (#3)}
\def\cma#1#2#3{{ Computers Math. Applic.} {\bf #1}, #2 (#3)}
\def\mc#1#2#3{{ Math. Compt.} {\bf #1}, #2 (#3)}
\def\apj#1#2#3{{ Astrophys. J.} {\bf #1}, #2 (#3)}
\def\apjs#1#2#3{{ Astrophys. J. Suppl.} {\bf #1}, #2 (#3)}
\def\acta#1#2#3{{ Acta Astronomica} {\bf #1}, #2 (#3)}
\def\apl#1#2#3{{Ann. Physik. (Leipzig)} {\bf #1}, #2 (#3)}
\def\anp#1#2#3{{Ann. Phys. } {\bf #1}, #2 (#3)}
\def\sa#1#2#3{{ Sov. Astro.} {\bf #1}, #2 (#3)}
\def\sia#1#2#3{{ SIAM J. Sci. Statist. Comput.} {\bf #1}, #2 (#3)}
\def\aa#1#2#3{{ Astron. Astrophys.} {\bf #1}, #2 (#3)}
\def\mnras#1#2#3{{ Mon. Not. R. astr. Soc.} {\bf #1}, #2 (#3)}
\def\npb#1#2#3{{ Nucl. Phys. B} {\bf #1}, #2 (#3)}
\def\prsla#1#2#3{{ Proc. R. Soc. London, Ser. A} {\bf #1}, #2 (#3)}
\def\jhep#1#2#3{{ JHEP} {\bf #1}, #2 (#3)}
\def\jgp#1#2#3{{ J.Geom.Phys.} {\bf #1}, #2 (#3)}

\def\nuc#1#2#3{{Nuovo Cimento B } {\bf #1}, #2 (#3)}
\def\ijmp#1#2#3{{Int. J. Mod. Phys. D} {\bf #1}, #2 (#3)}
\def\atmp#1#2#3{{Adv. Theor. Math. Phys.} {\bf #1}, #2 (#3)}
\def\ptps#1#2#3{{Prog. Theor. Phys. Suppl.} {\bf #1}, #2 (#3)}
\def\ptp#1#2#3{{Prog. Theor. Phys. } {\bf #1}, #2 (#3)}
\def\lmp#1#2#3{{Lett. Math. Phys. } {\bf #1}, #2 (#3)}
\def\mmj#1#2#3{{Mich. Math. j. } {\bf #1}, #2 (#3)}

%
\def\hepph#1#2{{ hep-ph }{\bf #1} (#2)}
\def\hepth#1#2{{ hep-th }{\bf #1} (#2)}
\def\grqc#1#2{{ gr-qc }{\bf #1} (#2)}
\def\ibid#1#2#3{{ {\it ibid.} }{\bf #1}, #2 (#3)}
%
\bibitem{chandra}
S.Chandrasekhar, {\it The Mathematical Theory of Black Holes}, Oxford University Press, Oxford (1992).

\bibitem{gib93}
G.W.Gibbons and A.R.Steif, \plb{314}{13}{1993}.
\bibitem{bar88}
R.Bartnik and J.McKinnon, \prl{61}{141}{1988}.

\bibitem{sak04}
I.Sakalli and M.Halilsoy, \prd{69}{124012}{2004}.
\bibitem{loh84}
D.Lohiya, \prd{30}{1194}{1984}.
\bibitem{fin00}
F.Finster, J.Smoller, and S.T.Yau, \atmp{4}{1231}{2000}.
\bibitem{findir}
F.Finster, J.Smoller, and S.T.Yau, \npb{584}{387}{2000},\\
F.Finster, J.Smoller, and S.T.Yau, \mmj{47}{199}{2000},\\
F.Finster, J.Smoller, and S.T.Yau, \cmp{205}{249}{1999},\\
F.Finster, J.Smoller, and S.T.Yau, \jmp{41}{2173}{2000}.

\bibitem{br1}
G.Silva-Ortigoza, \grg{33}{395}{2001}.
\bibitem{br2}
I.Sakalli, \grg{35}{1321}{2003}.

\bibitem{gib94}
G.W.Gibbons and A.R.Steif, \plb{320}{245}{1994}.
\bibitem{zec95}
A.Zecca, \jmp{37}{874}{1995}.
\bibitem{bel09}
F.Belgiorno and S.L.Cacciatori, \prd{79}{124024}{2009}.


\bibitem{pri72}
R.H.Price, \prd{5}{2419}{1972}.
\bibitem{gun94}
C.Gundlach, R.H.Price and J.Pullin, \prd{49}{883}{1994}.

\bibitem{pir1}
S.Hod and T.Piran, \prd{58}{024017}{1998}.
\bibitem{pir2}
S.Hod and T.Piran, \prd{58}{024018}{1998}.
\bibitem{pir3}
S.Hod and T.Piran, \prd{58}{024019}{1998}.

\bibitem{ja}
H.Koyama and A.Tomimatsu, \prd{63}{064032}{2001}.
\bibitem{ja1}
H.Koyama and A.Tomimatsu, \prd{64}{044014}{2001}.
\bibitem{mod01a}
R.Moderski and M.Rogatko, \prd{63}{084014}{2001}.
\bibitem{mod01b}
R.Moderski and M.Rogatko, \prd{64}{044024}{2001}.
\bibitem{rog07}
M.Rogatko, \prd{75}{104006}{2007}.
\bibitem{jin04}
J.L.Jing, \prd{70}{065004}{2004}.
\bibitem{jin05}
J.L.Jing, \prd{72}{027501}{2005}.
\bibitem{bur04}
L.M.Burko and G.Khanna, \prd{70}{044018}{2004}.
\bibitem{xhe06}
X.He and J.L.Jing, \npb{755}{313}{2006}.

\bibitem{gibrog08}
G.W.Gibbons and M.Rogatko, \prd{77}{044034}{2008}.

\bibitem{kon07}
R.A.Konoplya, A.Zhidenko, and C.Molina, \prd{75}{084004}{2007}.

\bibitem{br08}
G.W.Gibbons, M.Rogatko, and A.Szyplowska, \prd{77}{064024}{2008}.

\bibitem{vil}
A.Vilenkin and E.P.S.Shallard, {\it Cosmic Strings and Other Topological Defects},
Cambridge University Press, Cambridge (1994).

\bibitem{mon}
M.Bariola and A.Vilenkin, \prl{63}{341}{1989},\\
D.Harari and C.Lusto, \prd{42}{2626}{1990},\\
F.D.Mazziteli and C.Lusto, \ibid{43}{468}{1991},\\
H.Yu, \npb{430}{427}{1994}.
\bibitem{hyu02}
H.Yu, \prd{65}{087502}{2002}.
\bibitem{che05}
S.Chen and J.Jing, {\it Late-time Behaviour of a Coupled Scalar Field 
in Background of a Schwarzschild Black Hole with a Global Monopole},
\grqc{0511098}{2005}.
\bibitem{ary86}
M.Aryal, L.H.Ford, and A.Vilenkin, \prd{34}{2263}{1986}.
\bibitem{greg}
F.Dowker, R.Gregory, and J.Trashen, \prd{45}{2762}{1992},\\
R.Moderski and M.Rogatko, \ibid{57}{3449}{1998},\\
A.Achucarro, R.Gregory, and K.Kuijken, \ibid{52}{5729}{1995},\\
A.Chamblin, J.M.A.Ashbourn-Chamblin, R.Emparan, and A.Sorborger, \ibid{58}{12014}{1998},\\
F.Bonjour, R.Emparan, and R.Gregory, \ibid{59}{084022}{1999},\\
R.Moderski and M.Rogatko, \ibid{58}{124016}{1998},\\
C.Santos and R.Gregory, \ibid{61}{024006}{2000},\\
A.M.Ghezelbash and R.B.Mann, \ibid{65}{124022}{2002}.
\bibitem{mod99}
R.Moderski and M.Rogatko, \prd{60}{104040}{1999}.

\bibitem{qft}
C.Izykson and J.B.Zuber, {\it Quantum Field Theory}, McGrew Hill, New York (1985).

\bibitem{kim93}
M.Kim and M.K.Banerjee, \prc{48}{2035}{1993}.
\bibitem{cho75}
A.Chodos and C.B.Thorn, \prd{12}{2733}{1975}.

\bibitem{coo95}
F.Cooper, A.Khare, and Sukhatme, \prep{251}{267}{1995}.
\bibitem{and91}
A.Anderson and R.H.Price, \prd{43}{3147}{1991}.

\bibitem{cam96}
R.Camporesi and A.Higuchi, \jgp{20}{1}{1996}.

\bibitem{lea86}
E.W.Leaver, \prd{34}{384}{1986}.

\bibitem{abr70}
{\it Handbook of Mathematical Functions}, edited by M.Abramowitz and I.A.Stegun, (Dover, New York, 1970).

\bibitem{biz09}
P.Bizon, T.Chmaj, and A.Rostworowski, \cqg{26}{175006}{2009}.
\bibitem{oku08}
S.Okuzumi, K.Ioka, and M.Sakagami, \prd{77}{124018}{2008}.
\bibitem{linth}
Y.Sun and R.H.Price, \prd{38}{1040}{1988},\\
Y.Sun and R.H.Price, \prd{41}{2492}{1990},\\
H-P.Nollert and B.G.Schmidt, \prd{45}{2617}{1992},\\
N.Andersson, \prd{55}{468}{1997}.



\end{thebibliography}
\end{document}